\documentclass{article}%
\usepackage{amsmath}%
\usepackage{amsfonts}%
\usepackage{amssymb}%
\usepackage{graphicx}
\providecommand{\U}[1]{\protect\rule{.1in}{.1in}}

\begin{document}

\title{On the effect of wear on asperity heigth distributions, and the corresponding
effect in the mechanical response}
\author{M.Ciavarella\\Politecnico di BARI. \\V.le Gentile 182, 70125 Bari-Italy. \\Email mciava@poliba.it}
\maketitle

\begin{abstract}
Since the time of the original Greenwood \& Williamson paper, it was noticed
that abrasion and wear lead to possibly bimodal distribution of asperity
height distribution, with the upper tail of asperities following from the
characteristics of the process. Using a limit case solution due to Borucki for
the wear of an originally Gaussian distribution, it is shown here that the
tail is indeed always Gaussian, but with different equivalent parameters.
Therefore, if the wear process is light, one obtains a bimodal distribution
and both may affect the resulting contact mechanics behaviour. In this short
note, we illustrate just the main features of the problem. We conclude that it
is an oversimplification to consider surfaces Gaussian.

\end{abstract}

\section{Introduction}

From the times of the celebrated Greenwood and Williamson (1966) paper, we
know that surfaces that have been worn or abraded show a non-unique Gaussian
distribution. Indeed, Fig.6 of GW (adapted here as Fig.1) shows in a Gaussian
paper the distribution of summit heights in a surface of mild steel which had
been abraded and then slid against copper, which resembles a "bimodal
Gaussian".\ More precisely, Greenwood \& Wu 2001 returned to that very surface
and commented that Fig.6 of GW paper \textit{"shows an abraded surface which
is largely flat but where the heights of the upper 80\% may be regarded as
Gaussian. (We may note that `proper' statistical tests would not reveal this
fact)"}. 

\begin{center}
$%
\begin{array}
[c]{cc}%
{\includegraphics[
natheight=7.135600in,
natwidth=5.874700in,
height=3.5968in,
width=2.9654in
]%
{g66.jpg}%
}
&
\end{array}
$

Fig.1 - Height distribution of an abraded surface (adapted from Greenwood and
Williamson 1966).
\end{center}

On more careful consideration therefore, the surface had been largely worn
out, but this may not be always the case. Today, we can gather a lot more
insights into the process of wear or abrasion from the models developed, using
GW description of roughness, in the context of Chemical Mechanical\ Polishing
used for planarization of integrated circuits (Borucki, 2002, Borucki
\textit{et al.} 2004, Shi \& Ring 2010). Indeed, polishing causes high
asperities wearing faster than low asperities. Assume an original distribution
of asperities, $\phi_{0}\left(  z\right)  $ which is taken to be Gaussian (but
we postpone the term $\sqrt{2\pi}$ into the integrals for easy of notation)%
\begin{equation}
\phi_{0}=\frac{1}{\sigma}\exp\left[  -\frac{z^{2}}{2\sigma^{2}}\right]
\quad,\quad\overline{\phi}_{0}=\exp\left[  -\frac{z^{2}}{2}\right]
\end{equation}
where $\sigma$ is the rms amplitude of the summit heights.

Using Archard law for wear and the GW model, an equation similar to a
differential Hamilton-Jacobi equation was obtained for the distribution of
asperity heights which tends to develop very high peaks in the tail and indeed
in the case when the separation between the polishing pad and the surface is
kept constant at $z=c$, an analytical solution is developed with the method of
characteristics as (Borucki et al 2004, Shi \& Ring 2010)
\begin{align}
\phi\left(  z\right)    & =\phi_{0}\left(  z\right)  \quad,\quad z<c\\
\phi\left(  z\right)    & =\frac{w}{2\sqrt{z-c}}\left(  t+\frac{2}{w}%
\sqrt{z-c}\right)  \phi_{0}\left[  c+\frac{w^{2}}{4}\left(  t+\frac{2}{w}%
\sqrt{z-c}\right)  ^{2}\right]  \quad,\quad z>c\\
& =\frac{w}{2\sqrt{z-c}}\left(  t+\frac{2}{w}\sqrt{z-c}\right)  \phi
_{0}\left[  z+\frac{w^{2}}{4}t^{2}+\frac{w}{2}t\sqrt{z-c}\right]  \quad,\quad
z>c
\end{align}
where $w=c_{a}\frac{4}{2\pi}\frac{E^{\ast}}{\sqrt{R}}$, with $c_{a}=kV$ and
$k$ is the wear constant, while $V$ is sliding speed. Further, $E^{\ast}$ is
elastic modulus of materials, and $R$ the radius of asperity summits. As
Borucki et al (2004) clearly state:- "this solution develops an integrable
singularity at $z=d$ representing the portion of the pad-surface that has been
\textit{worn smooth} by the wafer. After a sufficiently long but finite time,
this singularity converges to a $\delta-$distribution with its amplitude
approaching a limiting value given by the fraction of the pad-surface
originally protruding above the wafer height". We need to take this carefully
in consideration when developing a contact mechanics model. Before we do that,
let us move to dimensionless form,
\begin{align}
\overline{\phi}\left(  \overline{z}\right)    & =\overline{\phi}_{0}%
\quad,\quad\overline{z}<\overline{c}\\
\overline{\phi}\left(  \overline{z}\right)    & =\left(  \frac{\overline{t}%
}{2\sqrt{\overline{z}-\overline{c}}}+1\right)  \overline{\phi}_{0}\left[
\overline{z}+\frac{\overline{t}^{2}}{4}+\overline{t}\sqrt{\overline
{z}-\overline{c}}\right]  \quad,\quad\overline{z}>\overline{c}%
\end{align}
where we have introduced a dimensionless time%
\begin{equation}
\overline{t}=wt/\sqrt{\sigma}%
\end{equation}

\begin{center}
\bigskip$%
\begin{array}
[c]{cc}%
{\includegraphics[
height=2.2857in,
width=3.659in
]%
{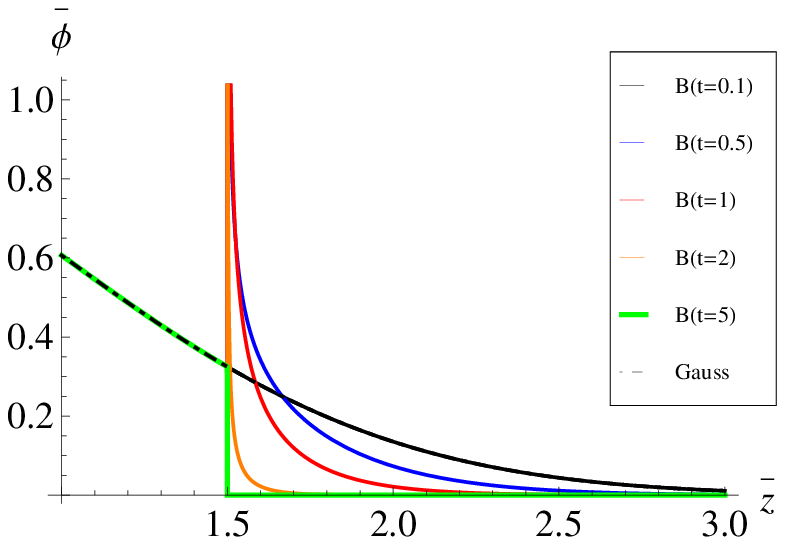}%
}
&
\end{array}
$

Fig.2. The distribution of asperity heights due to wear in Borucki's solution
$\overline{\phi}_{\operatorname{mod}}\left(  \overline{z},\overline
{c},\overline{t}\right)  $. $\overline{c}=1.5$, and $\overline{t}%
=0.1,0.5,1,2,5$. The delta function cannot be represented for obvious reasons.
\end{center}

This function is plotted in Fig.2 for representative dimensionless time
$\overline{t}=0.1,0.5,1,2,5$, and compared to the original Gaussian tail. 

We can integrate analytically the difference between the Borucki distribution,
and the original Gaussian one, to find the fraction of asperities worn out
\begin{equation}
F\left(  \overline{c},\overline{t}\right)  =-\int_{\overline{c}}^{\infty
}\left(  \overline{\phi}\left(  \overline{z}\right)  -\overline{\phi}%
_{0}\left(  \overline{z}\right)  \right)  d\overline{z}=-\sqrt{\frac{\pi}{2}%
}\left[  Erfc\left(  \frac{\overline{c}+\overline{t}^{2}/4}{\sqrt{2}}\right)
-Erfc\left(  \frac{\overline{c}}{\sqrt{2}}\right)  \right]
\end{equation}
which we shall assume, in the absence of a better hypothesis, all lye on the
$\overline{z}=\overline{c}$, and, perhaps an even stronger assumption,
maintain the same radius. Obviously if special experimental setups will permit
to have reasonably simple alternative assumptions, they could be readily incorporated.

Therefore, the so-modified Borucki distribution is
\begin{align}
\overline{\phi}_{\operatorname{mod}}\left(  \overline{z},\overline
{c},\overline{t}\right)    & =\overline{\phi}_{0}\quad,\quad\overline
{z}<\overline{c}\\
& =\left(  \frac{\overline{t}}{2\sqrt{\overline{z}-\overline{c}}}+1\right)
\overline{\phi}_{0}\left[  \overline{z}+\frac{\overline{t}^{2}}{4}%
+\overline{t}\sqrt{\overline{z}-\overline{c}}\right]  +F\left(  \overline
{c},\overline{t}\right)  \delta\left(  \overline{z}-\overline{c}\right)
\quad,\quad\overline{z}>\overline{c}%
\end{align}
where $\delta$ is the classical delta function.

\section{GW treatment}

By developing the usual GW treatment for number of asperities in contact, area
and load, integrating for the distribution of asperity heights, we obtain for
the compression $d=\left(  z_{s}-d_{0}\right)  $%
\begin{align}
n  & =\frac{N}{\sqrt{2\pi}}\int_{d_{0}}^{\infty}\phi\left(  z_{s}\right)
dz_{s}=\frac{N}{\sqrt{2\pi}}\int_{0}^{\infty}\overline{\phi}\left(
\overline{d}+\overline{d}_{0}\right)  d\overline{d}\\
\qquad A  & =N\frac{\pi}{\sqrt{2\pi}}R\int_{d_{0}}^{\infty}\left(  z_{s}%
-d_{0}\right)  \phi\left(  z_{s}\right)  dz_{s}=\pi NR\sigma\int_{0}^{\infty
}\overline{d}\overline{\phi}\left(  \overline{d}+\overline{d}_{0}\right)
d\overline{d}\\
P  & =\frac{4}{3\sqrt{2\pi}}E^{\ast}NR^{1/2}\sigma^{3/2}\int_{0}^{\infty
}\overline{d}^{3/2}\overline{\phi}\left(  \overline{d}+\overline{d}%
_{0}\right)  d\overline{d}%
\end{align}

We can define the Gaussian case integrals as
\[
I_{n}^{g}\left(  \overline{d}_{0}\right)  =\int_{0}^{\infty}\overline{d}%
^{n}\overline{\phi}_{0}\left(  \overline{d}+\overline{d}_{0}\right)
d\overline{d}%
\]
whereas the Borucki version "before the modification"
\[
I_{n}^{B}\left(  \overline{d}_{0},\overline{c},\overline{t}\right)  =\int
_{0}^{\infty}\overline{d}^{n}\overline{\phi}\left(  \overline{d}+\overline
{d}_{0},\overline{c},\overline{t}\right)  d\overline{d}%
\]
and after the modification, in considering the $\overline{\phi}%
_{\operatorname{mod}}$ functions, additional contributions lead to Heaviside
functions,
\begin{align}
I_{0\operatorname{mod}}^{B}\left(  \overline{d}_{0},\overline{c},\overline
{t}\right)    & =I_{0}^{B}\left(  \overline{d}_{0}\right)  +F\left(
\overline{c},\overline{t}\right)  H\left(  \overline{c}-\overline{d}%
_{0}\right)  \\
\qquad I_{1\operatorname{mod}}^{B}\left(  \overline{d}_{0},\overline
{c},\overline{t}\right)    & =I_{1}^{B}\left(  \overline{d}_{0}\right)
+F\left(  \overline{c},\overline{t}\right)  \left(  \overline{c}-\overline
{d}_{0}\right)  H\left(  \overline{c}-\overline{d}_{0}\right)  \\
I_{3/2\operatorname{mod}}^{B}\left(  \overline{d}_{0},\overline{c}%
,\overline{t}\right)    & =I_{3/2}^{B}\left(  \overline{d}_{0}\right)
+F\left(  \overline{c},\overline{t}\right)  \left(  \overline{c}-\overline
{d}_{0}\right)  ^{3/2}H\left(  \overline{c}-\overline{d}_{0}\right)
\end{align}

Further, writing
\[
P_{0}^{g}=\frac{4}{3}\frac{N}{\sqrt{2\pi}}ER^{1/2}\sigma^{3/2}%
\]
and
\[
A_{0}^{g}=\sqrt{\pi/2}\left(  NR\sigma\right)
\]
we can rewrite our final results as%
\begin{align}
n  & =\frac{N}{\sqrt{2\pi}}I_{0\operatorname{mod}}^{B}\left(  \overline{d}%
_{0},\overline{c},\overline{t}\right)  \\
\qquad A  & =A_{0}^{g}I_{1\operatorname{mod}}^{B}\left(  \overline{d}%
_{0},\overline{c},\overline{t}\right)  \\
P  & =P_{0}^{g}I_{3/2\operatorname{mod}}^{B}\left(  \overline{d}_{0}%
,\overline{c},\overline{t}\right)
\end{align}

\section{\bigskip Results}

The results\bigskip\ are quite as expected. Fig.3 plots the integrals
$I_{0\operatorname{mod}}^{B},I_{1\operatorname{mod}}^{B}%
,I_{3/2\operatorname{mod}}^{B}$ for the same case of Fig.2. Starting with the
integral giving the number of asperities in contact, it is clear that, at the
"jump" $\overline{c}=1.5$, we recover the number of asperities in the unworn
profile, by construction. Upon increasing the wear time, the tail is worn out
increasingly. Looking now at Fig.3b, this integral is proportional to the
contact area, and, as in Fig.3c, a change of slope results in moving across
the jump, but not an actual jump as in the number of asperities in Fig.3a.

\begin{center}
$%
\begin{array}
[c]{cc}%
{\includegraphics[
height=2.2857in,
width=3.659in
]%
{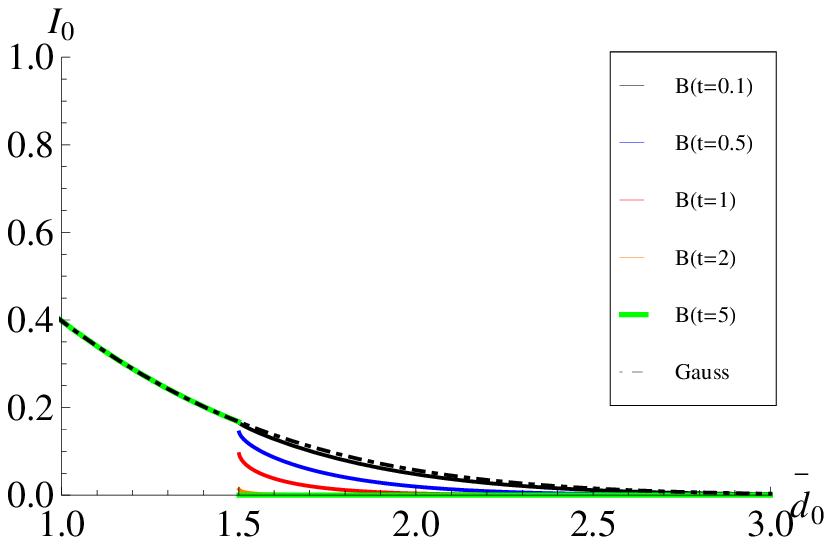}%
}
& (a)\\%
{\includegraphics[
height=2.2857in,
width=3.659in
]%
{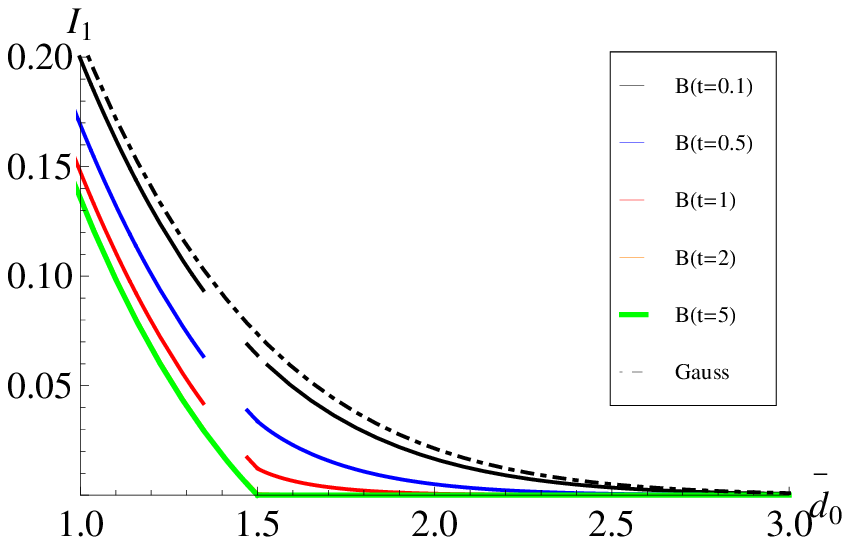}%
}
& (b)\\%
{\includegraphics[
height=2.2857in,
width=3.659in
]%
{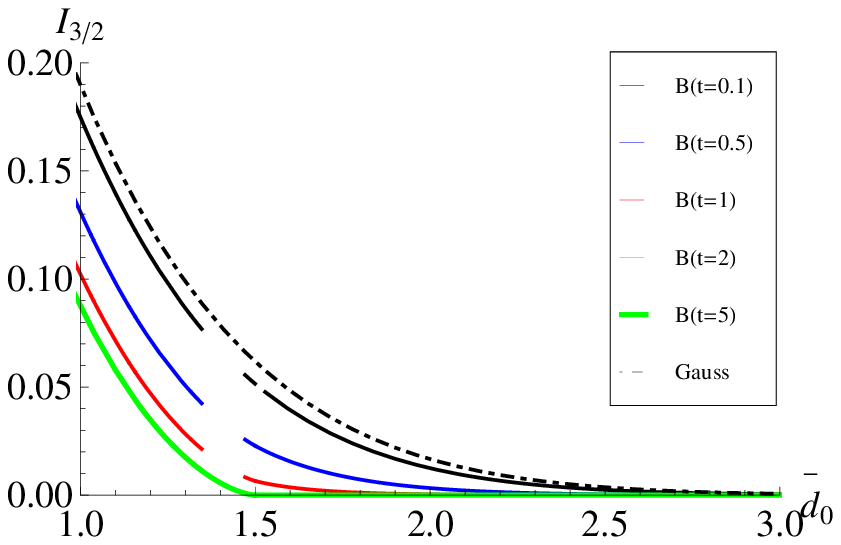}%
}
& (c)
\end{array}
$

Fig.3. The integrals $I_{0\operatorname{mod}}^{B},I_{1\operatorname{mod}}%
^{B},I_{3/2\operatorname{mod}}^{B}$ (a,b,c, respectively). $\overline{c}=1.5$,
and $\overline{t}=0.1,0.5,1,2,5$. 
\end{center}

We can define a mean pressure on the contact%
\begin{equation}
\overline{p}\left(  \overline{d}_{0},\overline{c},\overline{t}\right)
=\frac{P\left(  \overline{d}_{0},\overline{c},\overline{t}\right)  }{A\left(
\overline{d}_{0},\overline{c},\overline{t}\right)  }=\frac{4}{3\pi}%
E\frac{\sigma^{1/2}}{R^{1/2}}\frac{I_{3/2\operatorname{mod}}^{B}\left(
\overline{d}_{0},\overline{c},\overline{t}\right)  }{I_{1\operatorname{mod}%
}^{B}\left(  \overline{d}_{0},\overline{c},\overline{t}\right)  }%
\end{equation}
which for the unworn given Gaussian system, would be relatively constant with
separation. Fig.4 shows that the pressure is indeed more or less constant for
any system, despite the constant changes with time, decreasing as long as the
process continues -- this is mainly due to the fact that the "effective"
roughness is reduced in the tail of the distribution, because of wear. On the
other hand, if the contact moves to separations lower than $\overline{c}=1.5$,
then the mean pressure tends to restore rather quickly its original Gaussian
value --- it doesn't do immediately so because the asperities have now moved
to a lower height, and therefore, with respect to the unworn case, they are
compressed of a much lesser amount, resulting in less pressure. Obviously this
suggests that for the procedure to really work at constant separation
$\overline{c}$, the total load would need to decrease according to the model prediction.

\begin{center}
$%
\begin{array}
[c]{cc}%
{\includegraphics[
height=2.2857in,
width=3.659in
]%
{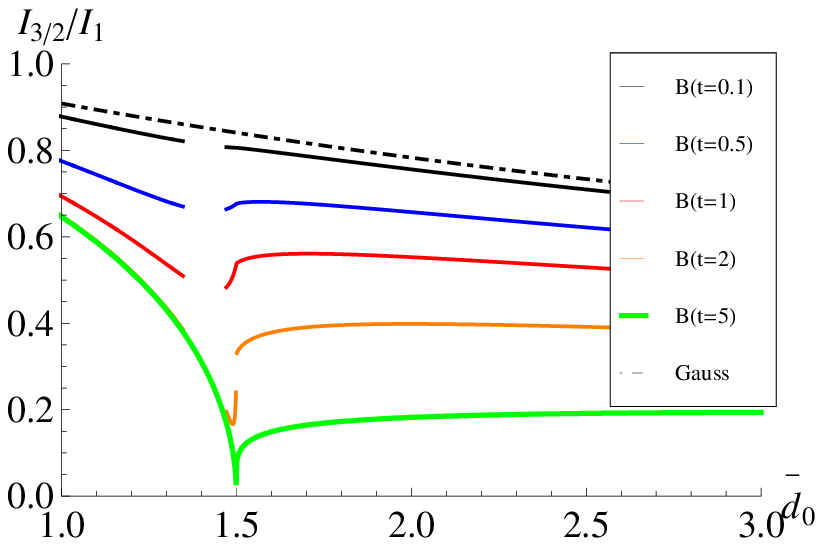}%
}
&
\end{array}
$

Fig.4. The ratio of the integrals $I_{3/2\operatorname{mod}}^{B}%
/I_{1\operatorname{mod}}^{B}$ which indicates mean pressure in the contact
areas. $\overline{c}=1.5$, and $\overline{t}=0.1,0.5,1,2,5$. 
\end{center}

\section{Conclusions}

We have shown in a simple case, that wear or abrasion leads to continuous
change of roughness parameters, which affect the response of the rough
contact. In cases when the process is selectively acting on part of the
asperity distribution, the effects are clearer, and a bimodal distribution
results. It may be therefore be an oversimplification to consider surfaces
Gaussian, as if coming simply from a single distribution.

\section{References}

Borucki, L. Mathematical modeling of polish rate decay in chemical-mechanical
polishing. J. Engng. Math. 43 (2002) 105--114

Borucki, L. J., Witelski, T., Please, C., Kramer, P. R., \& Schwendeman, D.
(2004). A theory of pad conditioning for chemical-mechanical polishing.
Journal of engineering mathematics, 50(1), 1-24.

Greenwood, J.A., Williamson, J.B.P., (1966). Contact of nominally flat
surfaces. Proc. R. Soc. London A295, 300---319.

Greenwood, J. A., \& Wu, J. J. (2001). Surface roughness and contact: an
apology. Meccanica, 36(6), 617-630.

Shi, H., \& Ring, T. A. (2010). Analytical solution for polish-rate decay in
chemical--mechanical polishing. Journal of Engineering Mathematics, 68(2), 207-211.

\end{document}